# Management of Network Slicing in 5G Radio Access Networks: Functional Framework and Information Models

R. Ferrús, O. Sallent, J. Pérez-Romero, R. Agustí

Universitat Politècnica de Catalunya (UPC)

[ferrus, sallent, jorperez, ramon]@tsc.upc.edu

*Abstract*— Network slicing is one of the key features for 5G networks to be able to accommodate the anticipated diversity of applications and business actors in a resource-efficient manner. While significant progress has already been achieved at 3GPP specifications level with regard to the system architectural and functional aspects for the realisation of network slicing in 5G networks, management solutions for the exploitation of this feature in the Next Generation Radio Access Network (NG-RAN) are still at a very incipient stage. In this context, this paper presents a framework for the management of network slicing in a NG-RAN infrastructure, identifying the functions, interfaces and information models that are necessary to enable the automation of the RAN slicing provisioning and management processes. Accordingly, a plausible information model intended to describe the manageable characteristics and behaviour of a *RAN slice* is developed and its applicability discussed in an illustrative neutral host provider scenario.

## Introduction

5G systems are being designed to support a wider range of applications and business models than previous generations due to the anticipated adoption of 5G technologies in multiple market segments (e.g. automotive, e-health, utilities, smart cities, agriculture, media and entertainment, high-tech manufacturing) and the consolidation of more flexible and cost-efficient service delivery models (e.g. neutral host network providers, Network as a Service, enterprise and private cellular networks). Through the support of network slicing, 5G systems are expected to become flexible and versatile network infrastructures where logical networks/partitions can be created (i.e. network slices) with appropriate isolation and optimized characteristics to serve a particular purpose or service category (e.g. applications with different access and/or functional requirements) or even individual customers (e.g. enterprises, third party service providers). This is especially relevant for the Radio Access Network (RAN), which is the most resource-demanding (and costliest) part of the mobile network and the most challenged by the support of network slicing [1].

System architecture and functional aspects to support network slicing in 5G Core Network (5GC) and Next Generation RAN (NG-RAN) have already been defined in the first release of the 5G normative specifications approved by 3GPP (e.g. network slice identifiers, procedures and functions for network slice selection, etc.) [2][3]. Moreover, implementation aspects of network slicing in the NG-RAN have been studied from multiple angles, ranging from virtualization techniques and programmable platforms with slice-aware traffic differentiation and protection mechanisms [4][5] to algorithms for dynamic resource sharing across slices [6]. In this respect, we analysed in [7] the RAN slicing problem in a multi-cell network in relation to how Radio Resource Management (RRM) functionalities can be used to properly share the radio resources and developed, in [8], a set of vendor-agnostic configuration descriptors intended to characterize the features, policies and resources to be put in place across the radio protocol layers of a NG-RAN node for the realization of concurrent *RAN slices*. On the other hand, management solutions necessary for the exploitation of network slicing capabilities in an automated and business agile manner are at a much more incipient stage, particularly for what concerns the NG-RAN. In this regard, a network slice lifecycle management solution for end-to-end automation across multiple resource domains is proposed in [9], including the RAN domain for completeness but not addressing it in details. More focused on a 5G RAN, [10] proposes the notion of an on-demand capacity broker that allows a RAN provider to allocate a portion of network capacity for a particular time period and [11] provide some insight on the need to extend current RAN management frameworks to support network slicing. At 3GPP level, standardisation work is on-going [12] to establish the general concepts, use cases and requirements for the management of network slicing across the whole set of the network functions specified in 3GPP (e.g. NG-RAN, 5GC and IMS) but without delving yet into the specifics of RAN slicing management.

In this context, aligned with current standardisation work at 3GPP and other industry fora (e.g. ETSI ISG NFV, Small Cell Forum), this paper develops a comprehensive functional framework for the management of network slicing in the NG-RAN, delineating the functions and information models that become necessary for the design, deployment and management operations of *RAN slices*. A central contribution of the paper is a plausible information model to represent the manageable characteristics of RAN slices. The remainder of the paper is organized as follows. First, a brief description of the NG-RAN architecture and its support for the network slicing feature as developed within 3GPP specifications is given. Second, the notion of *RAN Slice* is further elaborated and connected to the sort of resources found in a NG-RAN infrastructure. Next, the overall management architecture is presented and finally, an illustrative neutral host provider scenario is described to show the applicability of the proposed framework and information models.

THE NG-RAN: ARCHITECTURE AND RAN SLICES

*A. Functional architecture and support of network slicing*

From a functional perspective, the NG-RAN consists of two types of network functions (NFs)[1], each delivering the full radio access functionality to interact with the User Equipment (UE) over the radio interface: gNBs, using the 5G New Radio (NR) interface; and ng-eNBs, using an evolution of the LTE interface. Focusing on 5G NR access, as depicted in Fig. 1, gNBs are connected to the 5GC by means of NG interfaces [3] and may be interconnected with other gNBs and ng-eNBs through Xn interfaces. To introduce modularity and support different deployment options, 3GPP has also standardised the F1 interface that functionally splits a gNB into a gNB Central Unit (gNB-CU) for upper protocol layer processing and a gNB Distributed Unit (gNB-DU) for lower protocol layer processing. A single gNB, irrespective of whether it is split into gNB-CU/gNB-DU or not, handles the operation of one or more NR cells. Each NR cell, uniquely identified by a *cell ID*, is assigned with specific radio resources (i.e. RF carriers) which are operated under a common set of control channels (e.g. synchronisation, broadcast). The 5G NR interface is being designed with high flexibility, supporting Orthogonal Frequency-Division Multiplexing [OFDM]-based waveforms with different numerologies (e.g. different subcarrier spacing and cyclic prefix lengths) and adaptable time-frequency frame structures (e.g. selectable slot durations and dynamic assignment of DL/UL transmission direction). Moreover, the 5G NR interface allows for UEs served through the same NR cell to be instructed to receive or transmit using only a subset of the cell resource grid (feature known as bandwidth part contained operation). Ultimately, this flexibility of the NR interface allows UEs with diverse access types (e.g. enhanced Mobile Broadband [eMBB], massive Machine Type Communications [mMTC], Ultra Reliable Low Latency Communications [URLLC]) to be concurrently multiplexed over the same NR cell, as illustrated in Fig. 1.

From a service perspective, the overall 5G network (NG-RAN and 5GC) is designed to support a PDU Connectivity Service, i.e., a service that provides exchange of Protocol Data Units (PDUs) such as IPv4, IPv6, Ethernet or Unstructured data packets between a UE and an external data network reachable from the 5GC. The PDU Connectivity Service is realized via the establishment of one or multiple PDU sessions, which are the logical associations created between the 5GC and the UE to handle the data packet exchanges [2]. On this basis, the realization of network slicing relies on the principle that each PDU session is associated with a particular *Network Slice*.

---

[1] A Network Function (NF) is a processing function in a network which has defined functional behaviour and defined interfaces

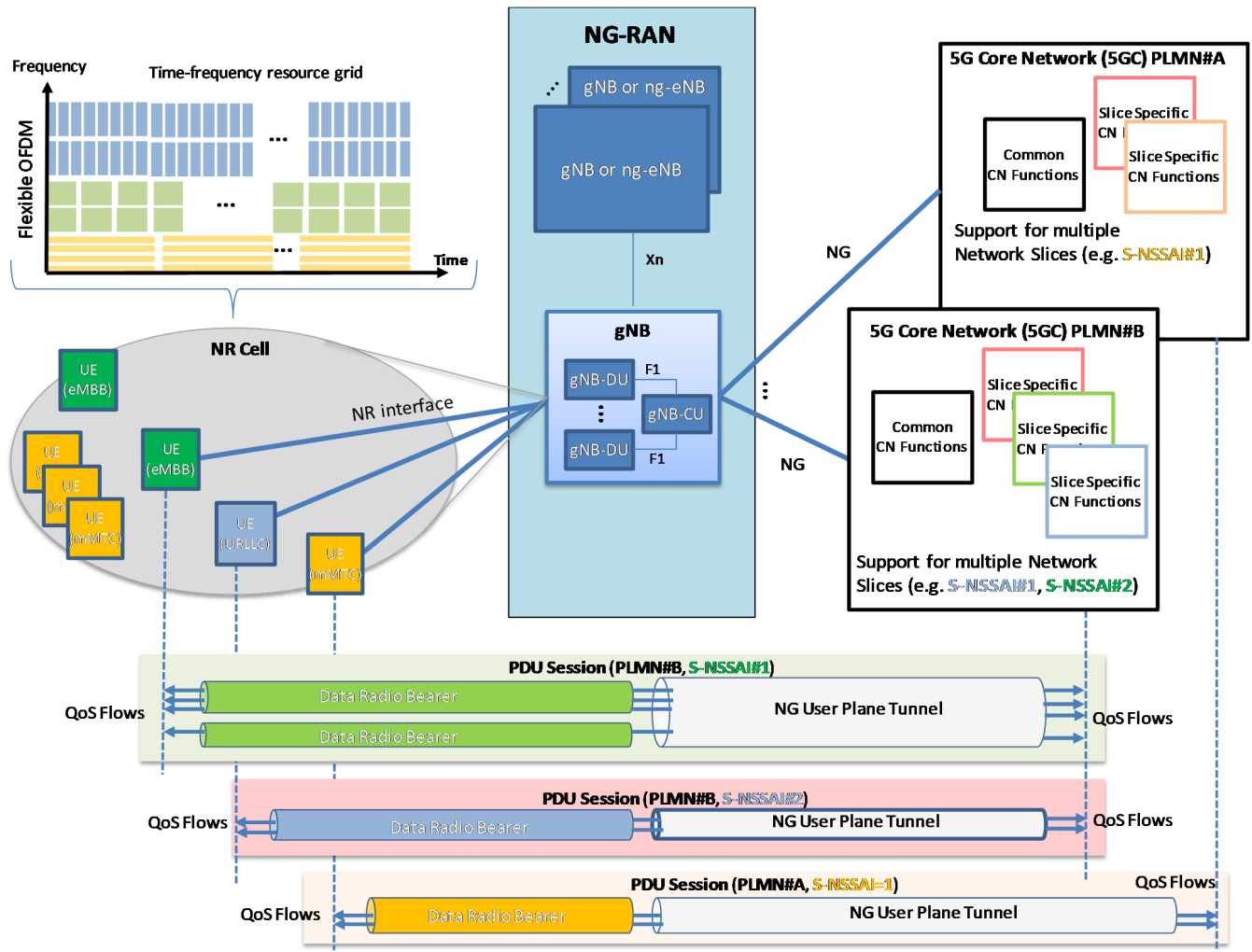

Fig. 1.- NG-RAN architecture with support for network slicing and RAN sharing

Indeed, a *Network Slice*, which is defined as a logical network that provides specific network capabilities and network characteristics, allows offering a differentiated network behaviour to UEs that are attached to the same 5G network (i.e. to the same 5GC, uniquely identified by a Public Land Mobile Network [PLMN] identity) but have PDU sessions associated with different delivered *Network Slices*. In addition to differentiated traffic treatment, *Network Slices* can also be used to serve different customers separately as per a given Service Level Agreement (SLA). Accordingly, a *Network Slice* is formally identified in 3GPP specifications [2] by a *Single Network Slice Selection Assistance Information* (S-NSSAI) identifier, which is unique within a PLMN and is comprised of a Slice/Service type (SST), denoting the expected network behaviour, and a Slice Differentiator (SD), differentiating amongst multiple network slices of the same Slice/Service type. Therefore, as illustrated in Fig. 1, each PDU session activated between a UE and a 5GC/PLMN network is associated with one and only one S-NSSAI so that the corresponding traffic flows (denoted as QoS flows in the 5G network) are handled according to whatever behaviour is pre-established for the selected S-NSSAI. The selection of the serving S-NSSAI is decided between the UE and 5GC based on e.g. subscription rights and communicated to the NG-RAN via signalling. Within the NG-RAN, the pre-established behaviour associated with the S-NSSAI can be then enforced by the proper handling of Data Radio Bearers (DRBs), which are the delivery services provided by the NG-RAN over the radio interface (e.g. specific scheduling rules and/or radio protocol stack configuration for the corresponding DRBs).

In addition to supporting multiple S-NSSAIs of a particular 5GC/PLMN, the NG-RAN could also serve multiple 5GC/PLMN networks by leveraging the sort of RAN sharing solutions introduced for legacy technologies such as 3GPP Multi-Operator Core Network (MOCN). Hence, gNBs could be connected to several 5GCs and the shared NR cells could broadcast information about the reachable 5GC/PLMN networks as well as support flexible access control mechanisms per PLMN/S-NSSAI (e.g. 5G Unified Access Control mechanisms).

*B. NG-RAN infrastructure and deployment of RAN Slices*

A particular NG-RAN behaviour, referred to as a *RAN Slice*, could be purposely designed to provide specific RAN capabilities (e.g. URLLC access) and network characteristics (e.g. coverage of an industrial park) to fulfil the access part requirements of the PDU connectivity service associated with a given S-NSSAI and PLMN. From a deployment perspective, the implementation of a *RAN Slice,* denoted as *RAN Slice Instance* (RSI), is likely to admit different possibilities as to how NG-RAN infrastructure functions and resources, including radio spectrum, are orchestrated and configured.

A NG-RAN infrastructure, illustrated in Fig. 2, is formed by a collection of physical sites (i.e. NG-RAN Points of Presence [PoPs]) including antenna installations (denoted as cell sites) and datacentre (DC) facilities, all interconnected by means of a fibre and/or wireless-based Transport Network (TN). As described in the previous section, the NG-RAN functionality for 5G NR access is delivered in the form of gNB NFs and/or a combination of gNB-CUs and gNB-DUs that will be deployed at cell sites or in the DCs. These NFs can be implemented as dedicated hardware appliances (referred to as Physical Network Functions [PNF] in ETSI Network Function Virtualization [NFV] terminology) or as Virtualized Network Functions (VNFs) running on a NFV Infrastructure (NFVI), which could be distributed among the DCs (the most part) but also extended to some of the cell sites (denoted *lightweight* NFVI in Fig. 2). The NG-RAN functionality most likely to be delivered as VNF is the gNB-CU as it does not embed the NR lower layers, though the virtualization of the full NR protocol stack is also a plausible scenario (e.g. gNB and gNB-DU provided as VNFs). Furthermore, for the sake of deployment flexibility, it is assumed that the RF systems at cell sites (Remote Radio Heads [RRHs]) are equipped with digital interfaces (based on CPRI protocol for instance) able to accommodate multiple digital baseband signal sources from co-located or remote gNB/gNB-DUs (V)NFs in a flexible manner over the supported frequency bands.

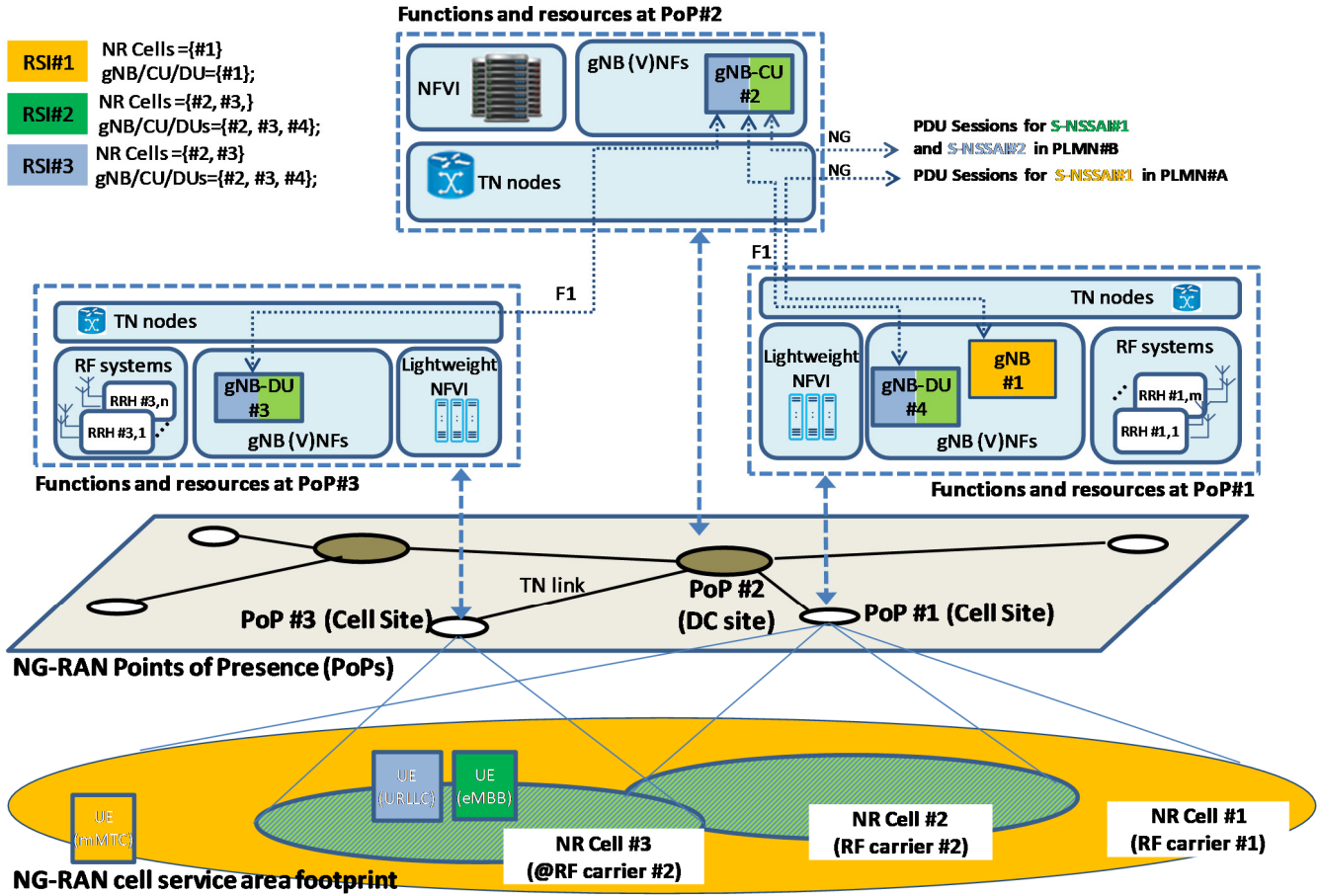

Fig. 2.- Illustration of a NG-RAN infrastructure and deployment of RAN Slice Instances (RSIs)

As an illustrative example, Fig. 2 shows the deployment of 3 different RSIs. RSI#1 is configured with dedicated spectrum resources (e.g. RF carrier #1), which are operated through a single dedicated cell covering a large area (e.g. NR Cell#1). This dedicated cell is handled by a dedicated gNB#1 VNF deployed at PoP#1. This could be an appropriate realization for full isolation in terms of spectrum and network functions between RSI#1 and the other RSIs. On the other hand, RSI#2 and RSI#3 are implemented over the same spectrum resources (e.g. RF carrier#2) and sharing a common set of cells (NR Cell #2 and #3), which consequently leads to sharing the same gNB functions realised in this case by means of gNB-CU#2 VNF deployed at PoP#2, gNB-DU#3 VNF at PoP#3 and gNB-DU#4 VNF at PoP#1. Notice that the implementation of RSI#2 and RSI#3 requires the support of network slicing features within gNB-CU#2, gNB-DU#3 gNB-DU#4 VNFs in order to properly distribute the shared cell resources among the two RSI according to e.g. the corresponding SLAs and desired level of isolation [7][8].

## MANAGEMENT OF RAN SLICES

The operator of a NG-RAN infrastructure shall be able to flexibly deploy and operate a number of concurrent RSIs that best serve its business needs. This includes the management capabilities for the creation, modification and termination of RSIs in an automated and business agile manner, allowing the operator to speed up deployment time as well as flexibly adapt the allocated resources and configuration of RSIs to keep track of the scenario dynamics (e.g. dynamic scaling of RSI capacity according to customer needs).

### A. *Functional framework for RAN slicing management*

Building upon the types of functions and resources of a NG-RAN infrastructure, Fig. 3 depicts a comprehensive architectural framework for the management of RSIs. The framework is consistent with 3GPP management reference model, which establishes a management architecture that separates between Element Management (EM) and Network Management (NM) layers integrated through standardized interfaces [13].

Following a bottom-up description approach, the NG-RAN NM layer sits on top of a collection of management systems that are specific to each of the function types composing the NG-RAN infrastructure (i.e. gNB functions, NFVI, RF systems, and TN nodes).

With regard to gNB functions, vendor-specific EM systems (EMSs) are assumed to be in place for detailed management. While EMSs commonly rely on proprietary interfaces and/or information models for the interaction with the managed functions, compatibility with 3GPP specifications mandates them to support a set of standardised interfaces, denoted collectively as Itf-N [13], for integration with the NM layer. Given that the NM layer is built upon vendor-neutral principles, Itf-N are object-oriented interfaces through which an information-centric view of the underlying network and services is provided. In particular, within the 3GPP management system, all functions and resources of the mobile network whose management is standardised are represented through *Network Resource Models* (NRMs) and *Interface* models. A NRM captures the manageable characteristics and behaviour of specific network functions and resources (e.g. NRM for CN functions, NRM for RAN functions, NRM for Self-Organizing Networks [SON] functionality). NRMs are specified as a collection of *Information Object Classes* (IOCs) together with their associations, attributes and operations. Likewise, *Interface* models specify the way that the information is accessed and manipulated (e.g.

bulk transfer). Current Itf-N specifications include support for Configuration Management (CM), Performance Management (PM) and Fault Management (FM) functional areas.

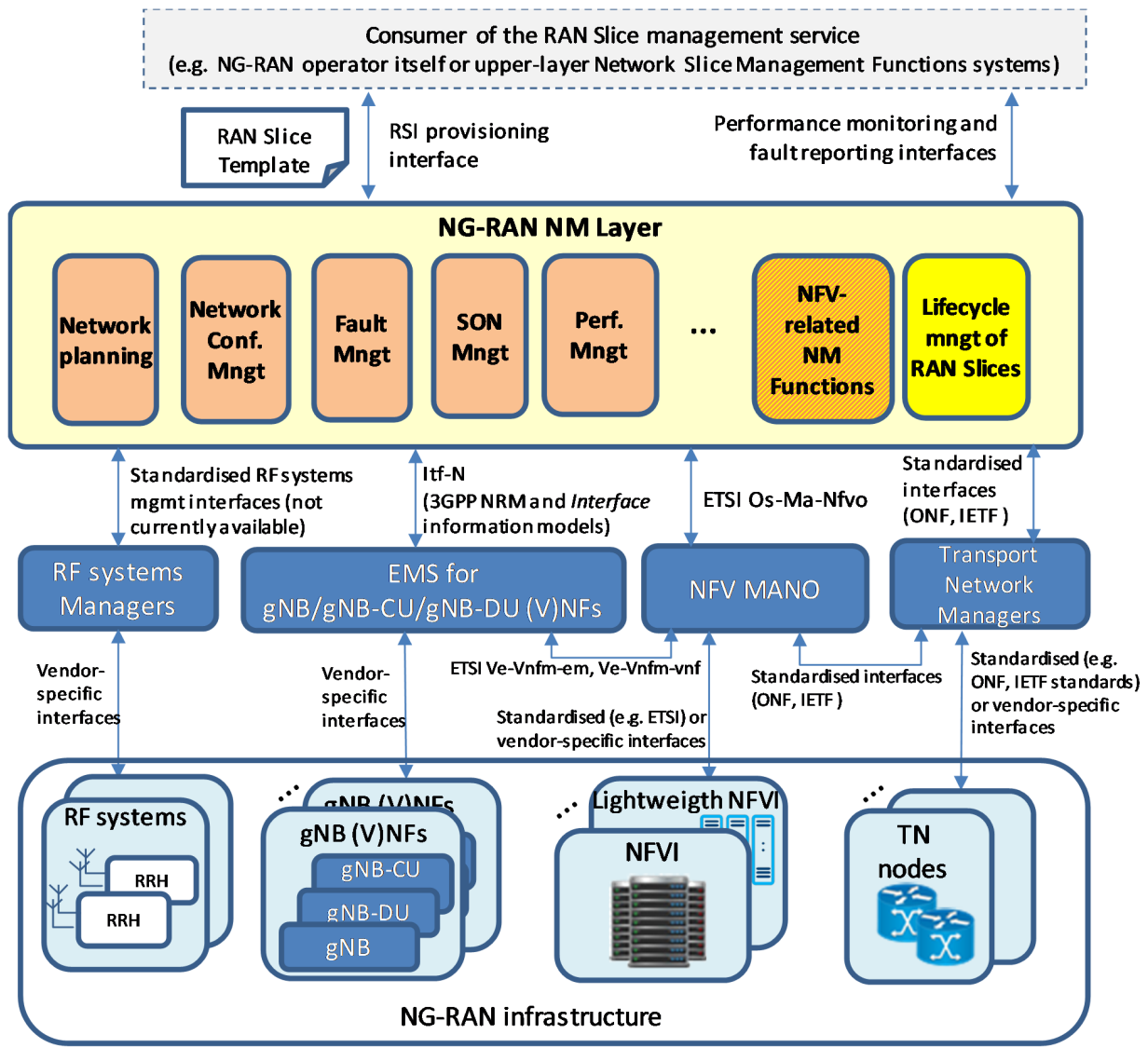

Fig. 3.- Architectural framework for the management of RAN slices

With regard to the NFVI resource-level aspects of the gNB functions delivered as VNF, ETSI NFV Management and Orchestration (MANO) compliant solutions are considered [14]. The ETSI NFV MANO allows the management of a distributed NFVI, encompassing the lifecycle management (creation, modification, termination) of individual VNFs as well as groups of interconnected VNFs and PNFs (i.e. Network Services [NS]). The integration of the ETSI NFV MANO solution and the NM and EM components relies on the adoption of the ETSI defined interfaces (e.g. Os-Ma-Nfvo interface between NM and ETSI NFV

MANO functions). In this respect, the NM layer can trigger the creation, modification or tear down of NSs and VNFs using ETSI standardised information models. In particular, ETSI NFV Network Service Descriptors (NSDs) can be used as deployment templates that describe, among others, the components of a NS (e.g. VNFs, PNFs, Virtual Links [VLs]), their relationship (e.g. forwarding graphs) and diverse deployment attributes (e.g. redundancy, scaling configuration).

With regard to the management of the TN connectivity, it is expected to benefit from the adoption of standardised SDN data models and technologies (e.g. ONF Transport API, YANG/NETCONF models for transport networks). Last but not least, RF systems are also assumed to be equipped with centralized management systems exposing functions for e.g. system configuration (e.g. operating region and bands, transmit power allocation) and signal distribution (e.g. signal sets creation and allocation) [15].

Moving up to the NM layer, this is traditionally composed of multiple and diverse management applications, such as network planning (e.g. determining the need to deploy new cells), network provisioning (e.g. configuration of cell parameters), centralised SON management (e.g. mobility robustness optimisation), network fault management (e.g. alarm correlation), network performance management (e.g. network monitoring) and so on. While still work in progress (3GPP TR 32.864), the introduction of virtualisation support in the NG-RAN requires to extend the NM layer with functions to support flexible installation, dimensioning, interconnection, healing and recovery of the part of the NG-RAN functionality implemented as VNFs. Of note is that neither the supported features nor the integration of such plethora of management applications within the NM layer is established by 3GPP, which only specifies the standardised interfaces and information models that are necessary for the interaction of the NM layer functionality with the rest of the management components.

In turn, support for RAN slicing management requires to further extend the NM layer with a new set of functions for the lifecycle management (LCM) of the RSIs (e.g. creation, modification, termination). In this respect, one challenging aspect is the translation between the *RAN Slice* requirements (e.g. business oriented requirements such as SLAs) and the specific design of a RSI in terms of allocated resources and configuration. This process may be efficiently solved through the involvement of network architect experts where new designs or re-designs of RAN slices occur infrequently. However, this may require automation through the specification of design templates (e.g. *RAN Slice Templates* in Fig. 3) and interfaces for the automated provisioning of RSIs in more dynamic scenarios (e.g. "RAN Slice as a Service" scenarios where RSI creation can be triggered from a business service order). Such provisioning interface would allow the NG-RAN operator (or an upper-level management function charged with the creation of an end-to-end network slice instance embedding the RSI in line with the on-going 3GPP work in [12]) to request the creation of a new RSI to satisfy specific requirements expressed through the *RAN Slice Templates* plus instance-specific configuration. In addition to automated provisioning, interfaces to expose information during the operation phase of a RSI for performance management (e.g. reporting measurement data for SLA compliance) and fault management (e.g. reporting upwards fault management data for operator or upper-level management system intervention) should also be provided by the NG-RAN NM layer. All in all, the realization of all these LCM functions and interfaces in the NM layer for RAN slicing management requires of new and/or extended NRM models that incorporate the necessary management objects for the configuration and operation of RSIs, as further developed in the following section.

*B. Information modelling for the characterisation of RAN slices*

A simplified view of a plausible NRM of the NG-RAN with support for RAN slice management is illustrated in Fig.4. The proposed model is based on the legacy E-UTRAN NRM model (3GPP TS 28.658), which we have adapted for NR cells and extended with two new IOCs used to characterise a RSI.

As depicted in Fig. 4, the root element of the whole construct is the *Subnetwork IOC*, which is used to group the set of managed entities (e.g. *ManagedElement IOC)* with the specific functionality to be managed (e.g. *gNBFunction* IOC) as seen over the Itf-N inteface. As long as a gNB NF could be configured to operate one or multiple NR cells, the corresponding *gNBFunction IOC* at NM level will be associated with one or multiple *NR Cell IOCs*, which are the central management elements that represent the properties of the deployed NR cells. More specifically, a *NRCell IOCs* contains attributes for specifying the frequency channel identifiers used by the cell, its transmission power, cell barring status and allowed access classes, etc. It also contains attributes for configuring RAN sharing aspects (e.g. list of PLMNs and management services such as PM/FM exposed per PLMN) and for establishing the relationships between cells and sector transmission equipment (e.g. Antennas, Tower Mounted Amplifiers [TMA]). Moreover, multiple additional IOCs can be associated to a *NRCell IOC* instance with regard to the management of diverse aspects (cell relationships, centralized SON, energy saving). On this basis, the functional representation of a managed RSI is formulated through the definition of the *cell slice* concept, which represents the particular realization of a RAN Slice into a given NR cell. This approach respects the cell-centric view of the current NRM model (i.e. most configuration as well as monitoring is primarily done at a cell basis) and easily introduces the concept of RAN slice as a group of cell slices. In more detail, as illustrated in Fig. 4, the two new IOCs introduced for that purpose are the *RANSlice IOC* and *CellSlice IOC*. The *RANSlice IOC* is conceived as the managed object class whose instances are used to represent the RSIs. The constituent elements of a *RANSlice IOC* instance are a collection of *CellSlice IOC* instances, where the *CellSlice IOC* represents the specific configuration of a NR cell to support the RSI.

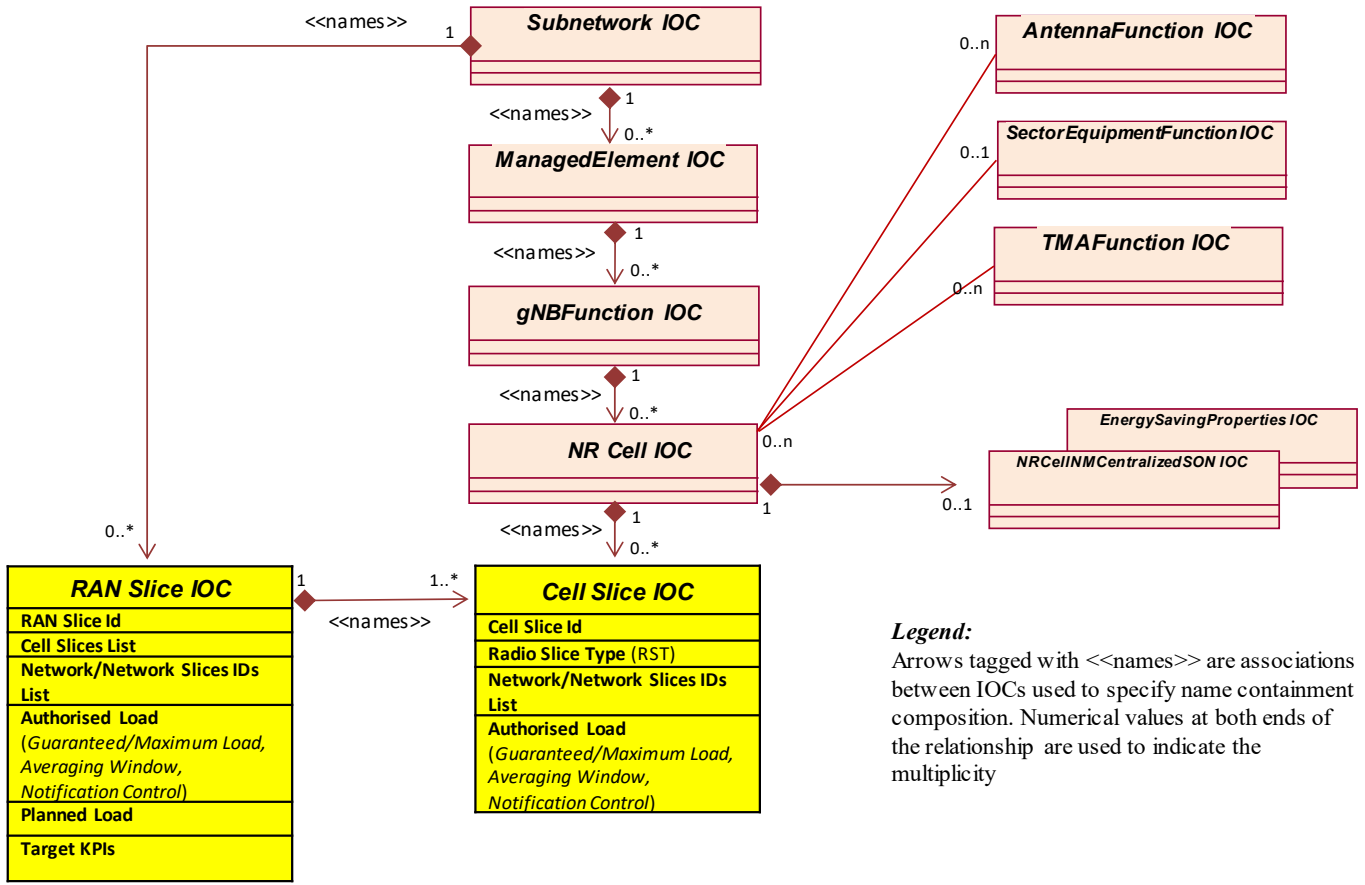

Fig. 4.- NG-RAN Network Resource Model (NRM) for RAN slicing management

The attributes proposed for the two new IOCs are detailed in Table 1. In contrast to having a detailed managerial view of the underlying RAN slicing capabilities supported in the underlying NG-RAN functions (e.g., detailed configuration of radio protocol capabilities and RRM parametrization for RAN slicing), the approach adopted here is based on keeping the RAN slice management simple at NM level. To that end, we consider that the detailed configuration of the NR cells for the realization of a particular behaviour is carried out at EM level, where more detailed vendor specific data models of the slicing capabilities supported in the (V)NFs can be used to pre-configure a set of radio protocol behaviours, denoted as Radio Slice Types (RSTs), as needed to properly support the access requirements of the set of SSTs (note that distinction between RST and SST is needed because there could be multiple SSTs with different global but common access behaviours that could be mapped to the same RST). On this basis, the set of RSTs would be exposed to the NM layer as a selectable list when creating/modifying a RSI. Moreover, in order to facilitate the automation of RSI provisioning with relatively low complexity, the provisioning of a RSI is sustained on the configuration of parameters that do not modify the cell footprint (e.g. do not impact on cell coverages, neighbour relations, mobility optimisations) but mainly impact on how the capacity of the cells is distributed among the RSIs. Hence, in addition to the RST attribute, the *CellSlice* IOC contains the list of supported network and network slice identifiers (i.e. *Network/Network Slices List* attribute) and the specification of the aggregated load restrictions and/or allowances for each QoS flow type or combination of QoS flow types served through the cell (i.e. *Authorised Load* [AL] attribute). In this way, the AL attribute, which is also included in the *RANSlice* IOC to allow establishing restrictions over the entire RSI, can be populated at NM level and communicated downwards through the Itf-N interface for enforcement. The details of the semantics of the AL attribute are given in Table 1. Finally, it is also foreseen that additional attributes may be incorporated into the *RANSlice* IOC to support the operation of other NM applications such as planning and SLA-compliance monitoring of the RSIs at NM level, as exemplified in Table I by means of the *PlannedLoad* and *TargetKPIs* attributes.

| New IOCs | Attributes |
|---|---|
| **Cell Slice IOC** | **Cell Slice Id**: It is used to identify the cell slice within a NR cell.<br>**Radio Slice Type (RST):** It is a scalar used to represent a differentiating radio protocol behaviour for UEs served through this cell slice. A pre-configured list of RSTs would be set up through the gNB EMS to support the requirements of the different standardized SSTs (e.g. eMBB, URLLC, mIoT) or any other operator-specific targeted behaviour.<br>**Network and Network Slice Identifiers List**: List of PLMN IDs and S-NSSAI served through the cell slice.<br>**Authorised Load (AL):** It is used to specify the aggregate capacity allowed or limited at cell level. Specified for the aggregate of each supported QoS flow type or combination of QoS flow types[1]:<br>    -*Guaranteed Load*: It denotes the minimum aggregate load that is provided across all activated QoS flows of the UEs served by the cell slice[2].<br>    -*Maximum Load*: It limits the aggregate load that can be expected to be provided across all activated QoS flows of the UEs served by the cell slice. Excess traffic may get blocked or discarded by a rate shaping function [3].<br>    -*Averaging Window*: It represents the duration over which the Guaranteed / Maximum Loads shall be calculated for compliance checking purposes.<br>    -*Notification Control*. It indicates whether notifications are requested from the EM to the NM layer when the Guaranteed Load cannot be fulfilled for the flow aggregate. It can take two values: *Enabled* (E)or *Disabled* (D). |
| **RAN Slice IOC** | **RAN Slice Id**: It is used to unambiguously identify a RAN Slice within a Subnetwork<br>**Cell Slices List**: It holds the list of the *Cell Slices* that compose the *RAN Slice*.<br>**Network and Network Slice Identifiers List:** Equivalent to the homonymous attribute within Cell Slice IOC but applicable at RAN slice level.<br>**Authorised Load (AL):** Equivalent to the homonymous attribute within Cell Slice IOC but applicable at RAN slice level.<br>**Planned Load:** It is used to characterise the load conditions of the RAN slice. This information is necessary for the proper dimensioning of resources allocated to the RAN slice and so establish the proper load restrictions and/or allowances.<br>**Target KPIs:** It is used to characterise a set of expected performance targets. This information is used in combination with the target / monitored load in the dimensioning process. Typical KPIs could be, for instance, average and minimum rates to be provided for Non-GBR QoS Flows. |

Notes:
(1) QoS Flow Types can be specified as pairs of 5G QoS Identifiers (5QI) and Allocation Retention Priority (ARP), which are the two parameters used to define the behaviour of QoS Flow in 5G systems.
(2) This is equivalent to the Guaranteed Flow Bit Rate (GFBR) parameter of individual QoS flows but applied to the load aggregate of all QoS flows activated of a given type or combination of types.

(3) This is equivalent to the Maximum Flow Bit Rate (MFBR) parameter of individual QoS flows but applied to the load aggregate of all QoS flows activated of a given type or combination of types.

Table 1. New Information Object Classes (IOCs) for the management of RAN Slices

APPLICABILITY EXAMPLE OF THE RAN SLICING MANAGEMENT FRAMEWORK

In order to gain insight into the proposed framework, let us consider a NG-RAN infrastructure like the one previously described in Fig. 2 where 3 RSIs are operational at a given stage. Such scenario could be the case of a neutral host provider that has deployed the NG-RAN infrastructure and offers RAN services to mobile network operators (MNOs) and other service providers (SP) in the form of RSIs. In particular, let us consider that RSI#1 is configured for mMTC services (e.g. utilities' metering application) offered through S-NSSAI#1 in a private 5GC/PLMN#A operated by a IoT SP and that, RSI#2 and RSI#3 are configured for, respectively, eMBB services offered through S-NSSAI#1 and Mission Critical (MC) services offered through S-NSSAI#2, in a public 5GC/PLMN#B operated by a MNO. In terms of traffic load, let us consider that RSI#1 provides access to multiple IoT devices spread in a large area and with very low aggregated throughput requirement. In contrast, RSI#2 and RSI#3 are assumed to serve higher bandwidth applications and traffic demand. From this configuration, in order to illustrate the dynamic management of the RAN slices, let us also assume that in a subsequent stage the following changes are introduced: (1) another RSI (RSI#4) is activated to serve a new network slice (S-NSSAI#1 in 5GC/PLMN#C) operated by another IoT SP that delivers low latency/low throughput applications for the connected car sector and (2) the capacity allocated to RSI#2 (commercial eMBB) is scaled up in order to meet a temporary hotspot situation.

On this basis, Fig. 5 shows, in the upper part, an illustration of the corresponding NRM and NSD models with the characterisation at NM level of the final configuration with the 4 described RSIs, and in the lower part, an illustrative view of the corresponding NG-RAN infrastructure with the deployed components and cell footprint. Boxes in red colour within the NRM/NSD models are used to identify the elements added/modified as a consequence of the changes carried out between the two stages. Likewise, the modifications of the NG-RAN infrastructure view in Fig. 5 with regard to the initial configuration depicted in Fig. 2 are also coloured in red.

Focusing first on the initial configuration, it can be observed that RSI#1 is realised through a single macro NR cell operating in B28 (i.e. 700 MHz band). This cell (NRCell#1) is configured with a bandwidth of 5 MHz (minimum cell bandwidth for NR) and a single cell slice (CellSlice#1) with RST=mMTC. The AL attribute in CellSlice#1 is not set in this initial configuration, so that no restrictions are imposed on RSI#1 traffic that could use the 100% of NRCell#1 capacity. The functions and resources utilised to support NRCell#1 are described in NSD#1, which contains the references to the gNB#1 VNF instantiated at PoP#1 as well as to the used RRHs and interconnection VLs. On the other hand, RSI#2 and RSI#3 are implemented by sharing 2 micro NR cells (NRCell#2 and #3 in Fig. 5), each one operating a channel bandwidth of 40 MHz in B42 (3.4-3.6 GHz). In this case, the functions and resources allocated to implement the two NR cells are described by NSD#2, which includes the references to the virtualised gNB-CU#2 instantiated at PoP#2, gNB-DU#3 at PoP#3 and gNB-DU#4 at PoP#1, together with the corresponding RRH PNFs and interconnection VLs. To properly split the radio capacity of NRCell#2 and #3, two cell slices are established per cell (one for RSI#2 and another for RSI#3). The cell slices associated with RSI#2 are configured with RST=eMBB and AL={70%, N/A, 10s, Enabled} for the whole set of used 5QI/ARPs. Likewise, the established configuration for RSI#3 is RST=eMBB and AL={30%, N/A, 10s, Enabled}. This configuration allows the MNO to guarantee a capacity distribution of 70% and 30% for, respectively, commercial MBB and MC services in congestion situations.

Focusing now on the final configuration, it can be seen that the new RSI#4 is activated as a new CellSlice#2 within NRCell#1. This CellSlice#2 is of type RST=URLCC due to the low latency requirement and capacity of NRCell#1 is now configured to be distributed equally between RSI#1 and RSI#4 by setting AL={50%, N/A, 10s, Enabled} in both CellSlice#1 and CellSlice#2 of NRCell#1. Moreover, it can be noted that no changes are required in NSD#1 to support RSI#4. On the other hand, to solve the capacity increase required in RSI#2, the illustrated decision corresponds to the activation of a new micro cell (NRCell#4) around PoP#3, which is considered to be the location of the NG-RAN infrastructure closest to the target hotspot area. To that end, a new gNB-DU#5 VNF is instantiated at PoP#3 in order to handle a new NRCell#4, as reflected in the modified NSD#2. In this example, NRCell#4 is assumed to be configured with an 80 MHz channel in B43 (3.6-3.8 GHz) and with a single cell slice (CellSlice#1) since the whole cell capacity is allocated to RSI#2. In this way, the aggregated capacity in the area around PoP#3 is significantly increased for RSI#2 as well as the possibility to considerably improve the offered peak rates of UE's connected to RSI#2 thanks to carrier aggregation between NRCell#4 and NRCell#3.

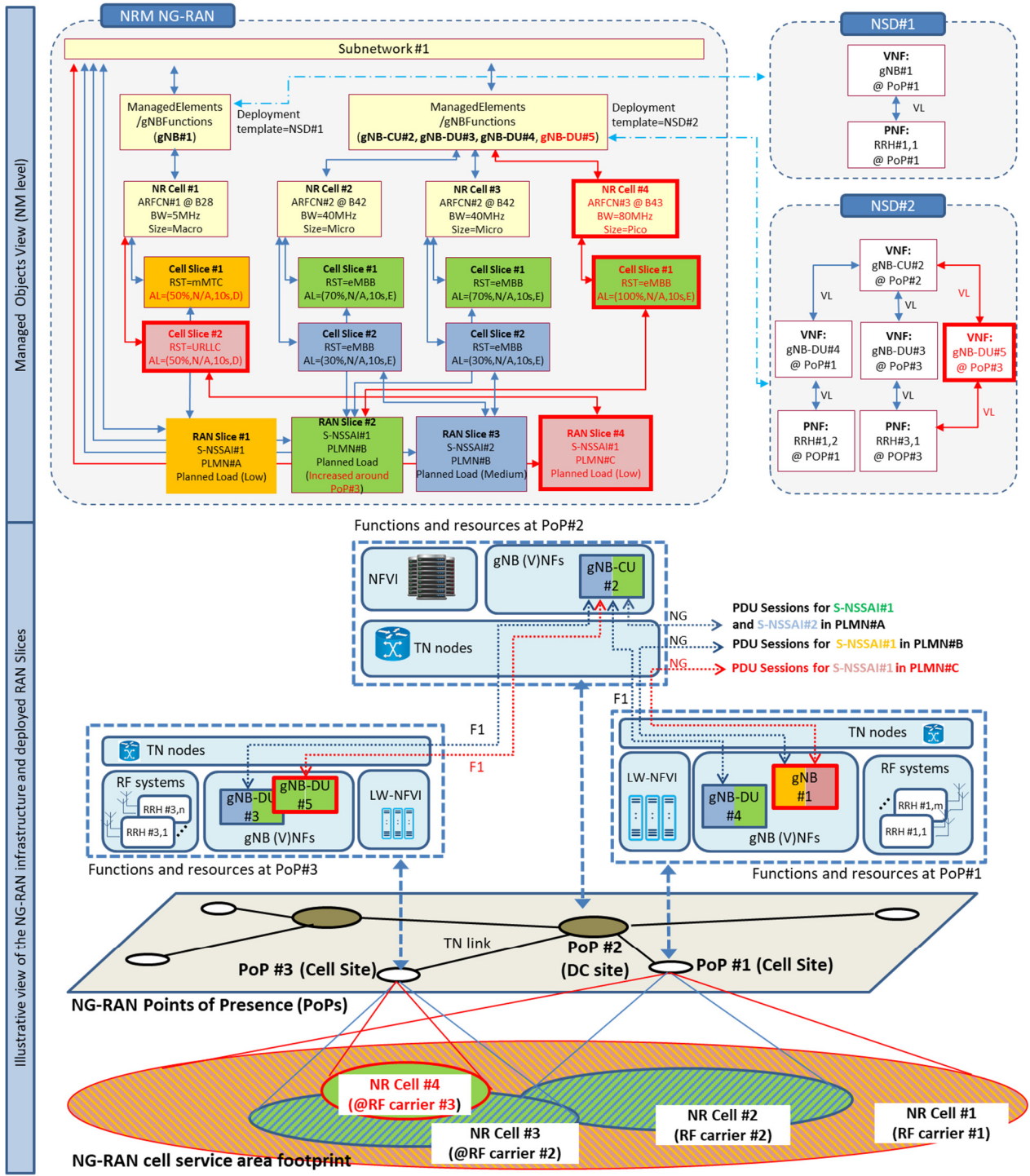

Fig. 5.- Modelling of the illustrative RAN slicing scenario through NRM and NSD information models

CONCLUDING REMARKS

The NG-RAN is expected to be a flexible and versatile access platform where multiple RAN slices can be dynamically created, thus radically transforming the legacy 3G/4G view of a static "one size fits all" access network. In order to achieve this goal, advanced management systems for network slicing are central components that need to be conceived, designed and implemented. In this respect, this paper has elaborated on a plausible functional framework for the realization of network slicing management in the NG-RAN, describing the new set of functions and managed object classes that should be incorporated into current network management frameworks. Special attention has been given to the necessary information models that become pivotal for articulating and assembling all of the required functionality. In this respect, the paper has proposed a functional representation of a managed RSI that respects cell-centric view of the NRM used in legacy systems, keeps the RAN slice management simple at NM level and facilitates the automation of RSI provisioning by acting on new IOC and attributes that mainly deal with network and network slice identifiers, RST selectors and thresholds for load guarantees/limitations to control the amount of traffic aggregate per RSI.